# New Superconducting Phase of $Li_x(C_6H_{16}N_2)_yFe_{2-z}Se_2$ with $T_c$ = 41 K Obtained through the Post-Annealing

Shohei Hosono, Takashi Noji, Takehiro Hatakeda, Takayuki Kawamata, Masatsune Kato, and Yoji Koike

*Department of Applied Physics, Tohoku University, 6-6-05 Aoba, Aramaki, Aoba-ku, Sendai 980-8579, Japan*

Post-annealing effects on the crystal structure and superconductivity of the lithium- and hexamethylenediamine (HMDA)-intercalated superconductor $Li_x(C_6H_{16}N_2)_yFe_{2-z}Se_2$ have been investigated. Through the post-annealing, a two-step reduction of the interlayer spacing between neighboring Fe layers, $d$, has been observed. It has been found that a new phase of $Li_x(C_6H_{16}N_2)_yFe_{2-z}Se_2$ with $d$ = 10.30(2) Å and $T_c$ = 41 K different from the as-intercalated phase is stabilized owing to the possible stable inclination of HMDA intercalated between FeSe layers. This result supports the domic relation between $T_c$ and $d$ in the FeSe-based intercalation superconductors. The reason why $T_c$ increases with a decrease in $d$ through the post-annealing is discussed.

## 1. Introduction

The compound FeSe has the simplest crystal structure among the iron-based superconductors composed of a stack of edge-sharing $FeSe_4$-tetrahedra layer by layers. Although the superconducting transition temperature, $T_c$, is only 8 K,[1] it dramatically increases up to ~ 37 K by the application of high pressure.[2,3] In single-layer FeSe films grown on $SrTiO_3$ substrates, recently, the opening of a possible superconducting gap has been observed at low temperatures below ~ 43 K and 65 K in the scanning tunneling microscopy/spectroscopy[4] and angle-resolved photoemission spectroscopy[5] measurements, respectively. Moreover, a superconducting transition has been detected at a high temperature above 100 K in the electrical transport measurements of the single-layer FeSe film.[6] These results suggest that $T_c$ of FeSe is very sensitive to the lattice and that FeSe has a great potential in achieving high $T_c$'s. Through the intercalation of potassium[7] and sodium[8] into FeSe, in fact, the interlayer spacing between neighboring Fe layers, $d$, has increased from 5.52 Å to 7.02 Å and 6.83 Å, so that $T_c$ has increased up to ~ 31 K and ~ 37 K, respectively, while $T_c$ of lithium-intercalated FeSe without increase in $d$ has remained 8 K.[9] Further increase in $d$ has been attained through the co-intercalation of alkali or alkaline-earth metals and ammonia, so that $d$ and $T_c$ have increased to 7 - 11 Å and 31 - 46 K in $A_x(NH_3)_yFe_{2-z}Se_2$ ($A$: alkali or alkaline-earth metals), respectively.[8,10-14] Following ammonia, several kinds of organic molecules have also been reported to be intercalated into FeSe together with alkali or alkaline-earth metals. In pyridine-intercalated $Li_x(C_5H_5N)_yFe_{2-z}Se_2$, $d$ and $T_c$ have increased to 8.03 - 11.55 Å and 45 K, respectively.[15] In ethylenediamine-intercalated $A_x(C_2H_8N_2)_yFe_{2-z}Se_2$ ($A$ = Li, Na), $d$ and $T_c$ have increased to 10.4 - 11.0 Å and 45 K, respectively.[16,17] Collecting these results of the FeSe-based intercalation superconductors, it has been found that $T_c$ tends to increase with increasing $d$ and to be saturated at ~ 45 K for $d \geq 9$ Å and that the carrier doping of intercalants into the conducting FeSe layer seems not to affect $T_c$ so much.[17] This tendency has been interpreted by Guterding $et\ al.$[18] in terms of the electronic structure calculated with the density functional theory. That is, the electronic structure becomes more two-dimensional with increasing $d$ and is perfectly two-dimensional at $d$ = 8 - 10 Å, leading to the saturation of $T_c$ at $d \geq 8$ Å. Within the random-phase-approximation spin-fluctuation approach, moreover, they have predicted that $T_c$'s of the FeSe-based intercalation superconductors with the $S_\pm$ symmetry increase not through the further increase in $d$ but through the electron doping owing to the increase of the density of states at the Fermi level at $d \geq 8$ Å. In linear-diamine-intercalated $A_x(C_nH_{2n+4}N_2)_yFe_{2-z}Se_2$ ($A$ = Na, Sr ; $n$ = 0, 2, 3) with $d$ = 8.7 - 11.4 Å, in fact, $T_c$ is not so dependent on $d$ and $T_c$'s of heavily electron-doped Na-intercalated compounds are 41 - 46 K and higher than those (34 - 38 K) of lightly doped Sr-intercalated ones.[19]

Recently, we have succeeded in synthesizing lithium- and hexemethylenediamine (HMDA)-intercalated $Li_x(C_6H_{16}N_2)_yFe_{2-z}Se_2$ whose $d$ value is as large as 16.225(5) Å and the largest among those of the FeSe-based intercalation superconductors. The $T_c$ is 38 K and a little lower than the saturated value of ~ 45 K.

In the present work, we have investigated post-annealing effects on the crystal structure and superconductivity in $Li_x(C_6H_{16}N_2)_yFe_{2-z}Se_2$. A new phase of $Li_x(C_6H_{16}N_2)_yFe_{2-z}Se_2$ with $d$ = 10.30(2) Å and $T_c$ = 41 K different from the as-intercalated phase has been discovered, and the dependence of $T_c$ on $d$ in the FeSe-based intercalation superconductors is discussed.

## 2. Experimental

Polycrystalline host samples of FeSe were prepared by the high-temperature solid-state reaction method as described in Ref. 8). That is, iron powder and selenium grains, which were weighted in the molar ratio of Fe : Se = 1.02 : 1, were mixed, put into an alumina crucible and sealed in an evacuated quartz tube. This was heated to 1027°C for 30 h and then annealed at 400°C for 50 h, followed by furnace-cooling. The obtained ingot of FeSe was pulverized into powder to be used for the intercalation. Lithium- and HMDA-intercalated samples of $Li_x(C_6H_{16}N_2)_yFe_{2-z}Se_2$ were prepared as follow.[20] First, 0.8 M solution of lithium metal dissolved in HMDA was prepared in a beaker and the powdery FeSe was put into the solution in the molar ratio of Li : FeSe = 1 : 2. The reaction was carried out at 45°C for 2 days. The separation of the product from residual HMDA was easily performed by the solidification of residual HMDA at the top cap of the beaker, keeping the temperature of the top cap of the beaker below the melting point of HMDA (42°C). All the processes were carried out in an argon-filled glove box. Post-annealing of as-intercalated samples was performed at 100 - 250°C for 60 h in an evacuated glass tube. Both FeSe and the intercalated samples were characterized by the powder x-ray diffraction using CuK$_\alpha$ radiation. For the as-intercalated and post-annealed samples, an airtight sample-holder was used. The diffraction patterns were analyzed using RIETAN-FP.[21] The chemical composition was determined by the inductively coupled plasma optical emission spectrometry (ICP-OES). Thermogravimetric (TG) measurements were performed in flowing gas of argon, using a commercial analyzer (SII Nano Technology Inc., EXSTAR DSC7020). In order to detect the superconducting transition, not only magnetic susceptibility, $\chi$, measurements using a superconducting quantum interference device (SQUID) magnetometer (Quantum Design, MPMS) but also electrical resistivity, $\rho$, measurements by the standard dc four-probe method were carried out. For the $\rho$ measurements, the as-intercalated powdery sample was pressed into a pellet and sintered at 150°C for 12 h in vacuum (~ $10^{-1}$ Pa) using an oil-rotary pump.

## 3. Results and Discussion

Figure 1 shows powder x-ray diffraction patterns of the host compound FeSe, the as-intercalated sample and samples post-annealed at various temperatures for 60 h. It is found that the as-intercalated sample is mainly made up of $Li_x(C_6H_{16}N_2)_yFe_{2-z}Se_2$ with the ThCr$_2$Si$_2$-type structure and that there remain small regions of non-intercalated FeSe with the PbO-type structure in the as-intercalated sample. As for the post-annealed samples, it is found that diffraction peaks of $Li_x(C_6H_{16}N_2)_yFe_{2-z}Se_2$ remain in samples post-annealed below 200°C and that only peaks due to FeSe are observed in the sample post-annealed at 250°C, suggesting the perfect deintercalation of HMDA at 250°C. Interestingly, both 002 and 004 peaks of $Li_x(C_6H_{16}N_2)_yFe_{2-z}Se_2$ shift toward the high-angle side through the post-annealing, meaning that the $c$-axis length, namely, the $d$ value decreases with increasing post-annealing temperature. Both the lattice constant $c$ and $d$ of $Li_x(C_6H_{16}N_2)_yFe_{2-z}Se_2$ are listed in Table I. Here, the $d$ value is given by the half of the $c$-axis length, for the unit cell of $Li_x(C_6H_{16}N_2)_yFe_{2-z}Se_2$ includes two Fe layers. It is of interest that the $d$ value decreases not continuously but discretely from 16.17(3) Å (Phase I) in the as-intercalated sample to ~ 14 Å (Phase II) and ~ 10 Å (Phase III) with increasing post-annealing temperature, indicating the appearance of phases different from the as-intercalated one.

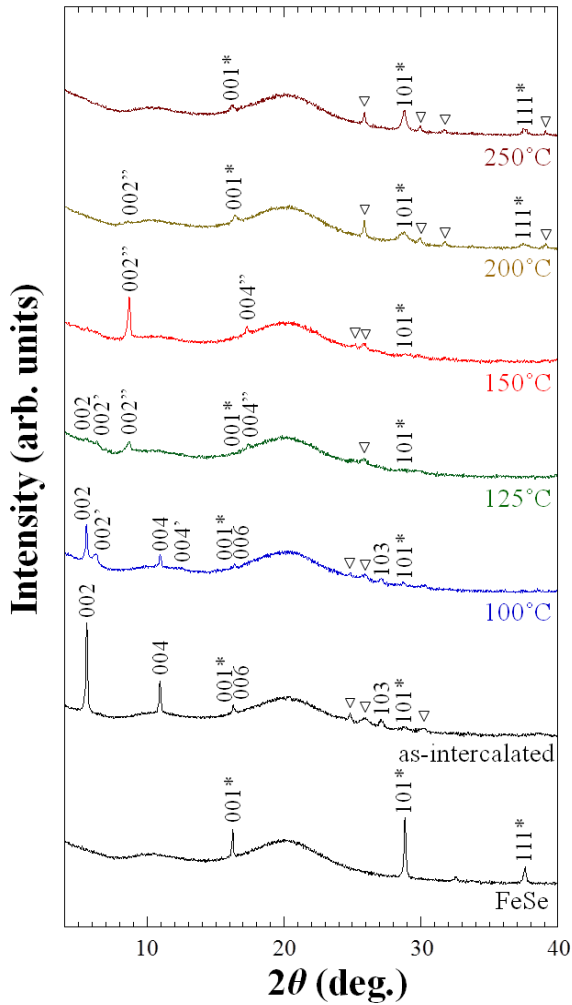

Fig. 1. (Color online) Powder x-ray diffraction patterns of the host compound FeSe and the as-intercalated and post-annealed (100 - 250°C, 60 h) samples consisting of $Li_x(C_6H_{16}N_2)_yFe_{2-z}Se_2$ (Phase I) and FeSe, using $CuK_\alpha$ radiation. Indices without and with an asterisk are based on the $ThCr_2Si_2$-type and PbO-type structures, respectively. Indices with single and double quotation marks represent Phase II and Phase III different from the as-intercalated Phase I, respectively. Peaks marked by ▽ are due to unknown compounds. The broad peak around $2\theta$ = 20° is due to the airtight sample-holder.

Table I. Lattice constant $c$ (in Å) and the interlayer spacing between neighboring Fe layers, $d$ (in Å), of $Li_x(C_6H_{16}N_2)_yFe_{2-z}Se_2$ in the as-intercalated and post-annealed samples. It is noted that $d$ is given by the half of $c$.

| Annealing Temperature (°C) | $c$ | | | $d$ | | |
|---|---|---|---|---|---|---|
| | Phase I | Phase II | Phase III | Phase I | Phase II | Phase III |
| As-intercalated | 32.33 (6) | | | 16.17 (3) | | |
| 100 | 32.42 (7) | 28.17 (6) | | 16.21 (4) | 14.09 (3) | |
| 125 | 32.62 (5) | 28.00 (2) | 20.44 (1) | 16.31 (3) | 14.00 (1) | 10.22 (1) |
| 150 | | | 20.59 (4) | | | 10.30 (2) |
| 200 | | | 20.68 (1) | | | 10.34 (1) |

The composition of the as-intercalated sample has been estimated by ICP-OES as $Li_{1.25}(C_6H_{16}N_2)_{0.64}Fe_{2.05}Se_2$. However, it is noted that the composition should be somewhat different from that of the region actually including lithium and HMDA, because there remain non-intercalated regions of FeSe in the sample. The ICP-OES analysis has also revealed that the contents of Li, Fe and Se do not change so much through the post-annealing up to 250°C.

Figure 2 shows the typical TG curve on heating up to 900°C at the rate of 1°C/min for the as-intercalated sample. As reported in our previous paper,[20] the mass loss below ~ 250°C is in good correspondence with the weight percent of HMDA in the as-intercalated sample estimated by ICP-OES. Therefore, it is concluded that HMDA in the as-intercalated sample deintercalates with increasing temperature, so that it almost perfectly deintercalates above ~ 250°C. This is in good correspondence with the powder x-ray diffraction result shown in Fig. 1.

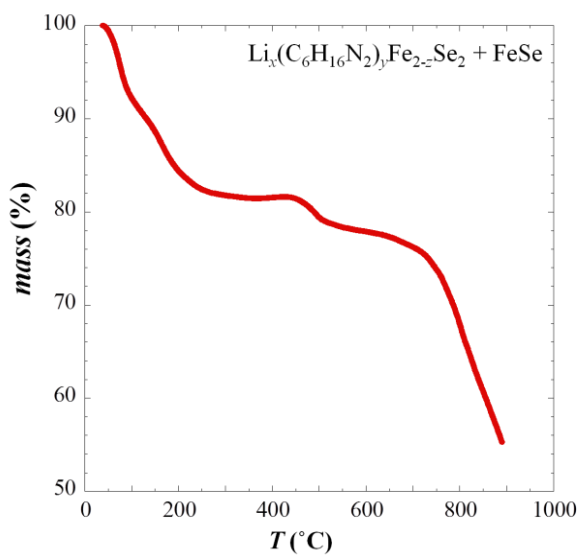

Fig. 2. (Color online) Thermogravimetric (TG) curve on heating at the rate of 1°C/min for the as-intercalated sample consisting of $Li_x(C_6H_{16}N_2)_yFe_{2-z}Se_2$ and FeSe.

Considering the powder x-ray diffraction, ICP-OES and TG results, crystal structures of the as-intercalated sample (Phase I) and post-annealed samples (Phase II and Phase III) are speculated as follows. First, the crystal structure of Phase I is speculated as shown in Fig. 3(a), because the $d$ value of 16.17(3) Å is just large enough for linear molecules of HMDA of 10.40 Å in length to be intercalated perpendicularly to the FeSe layers. This is similar to linear molecules of ethylenediamine of 5.37 Å in length in $A_x(C_2H_8N_2)_yFe_{2-z}Se_2$.[16] $Li^+$ ions seem to be located near both edges of HMDA, because the content of Li is roughly twice larger than that of HMDA according to the ICP-OES result, and because $Li^+$ ions will favor N and Se with comparatively large electronegativity. Next, crystal structures of Phase II and Phase III are speculated as shown in Figs. 3(b) and (c), respectively. As mentioned above, HMDA deintercalates so much with increasing post-annealing temperature, so that HMDA cannot keep the large $d$ to be forced to incline. Taking into account the discrete values of $d$ and the length of HMDA, the angle between the intercalated HMDA and the FeSe layers is calculated to be ~ 60° and ~ 30° in Phase II and Phase III, respectively. To the surprise, the positions of two N atoms of one HMDA molecule are displaced by approximately one and two unit-cells with each other, respectively, as illustrated in Figs. 3(b) and (c). This is plausible, because the length of HMDA is too short for the structure where the positions of two N atoms of one HMDA molecule slip each other by three unit-cells to be stabilized and moreover no x-ray diffraction peaks due to the structure have been observed. To confirm these speculations, the detailed analysis of the crystal structures is necessary.

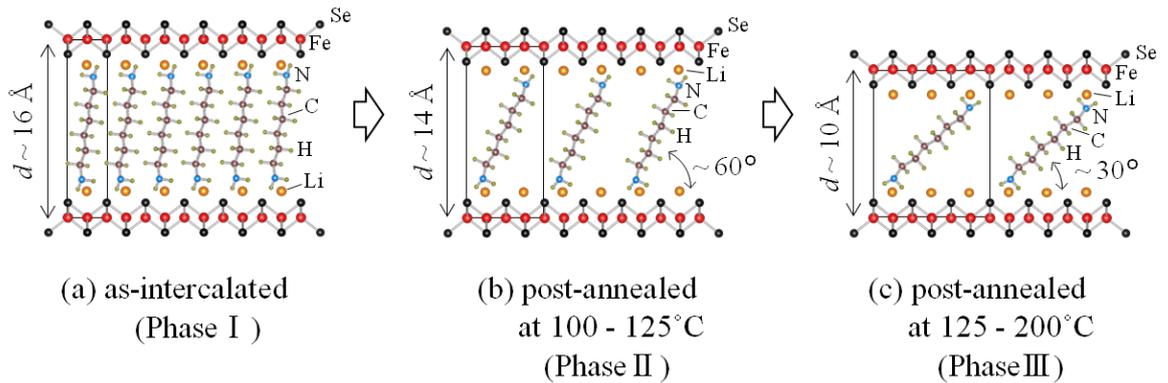

Fig. 3. (Color online) Schematic views of crystal structures of (a) the as-intercalated sample of $Li_x(C_6H_{16}N_2)_yFe_{2-z}Se_2$ with $d$ ~ 16 Å (Phase I), (b) samples post-annealed at 100 - 125°C with $d$ ~ 14 Å (Phase II) and (c) samples post-annealed at 125 - 200°C with $d$ ~ 10 Å (Phase III). Solid lines represent halves of the respective unit-cells.

Figure 4 shows the temperature dependence of $\chi$ in a magnetic field of 10 Oe on zero-field cooling (ZFC) and on field cooling (FC) for the as-intercalated sample and samples post-annealed at various temperatures for 60 h. As reported in our previous paper,[20] the first and second transitions at 38 K and 8 K in the as-intercalated sample are due to bulk superconductivity of Phase I and non-intercalated FeSe, respectively. It is found that $T_c$ of Li$_x$(C$_6$H$_{16}$N$_2$)$_y$Fe$_{2-z}$Se$_2$ increases to 41 K for samples post-annealed at 100 - 150°C and that $T_c$ is clearly 41 K in the sample post-annealed at 150°C of Li$_x$(C$_6$H$_{16}$N$_2$)$_y$Fe$_{2-z}$Se$_2$ only including Phase III, taking into account the powder x-ray diffraction result shown in Fig. 1.

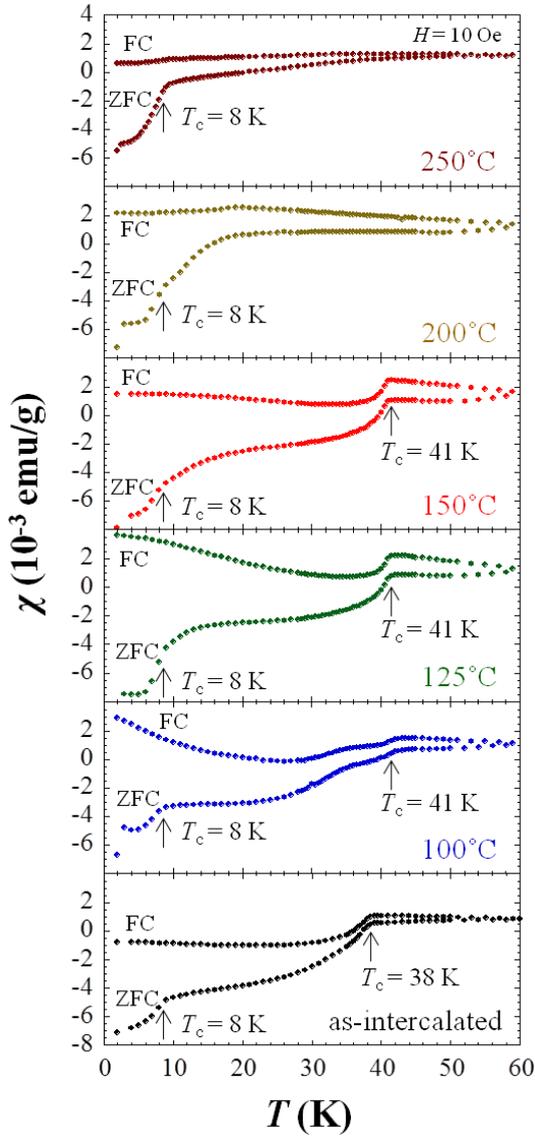

Fig. 4. (Color online) Temperature dependence of the magnetic susceptibility, $\chi$, in a magnetic field of 10 Oe on zero-field cooling (ZFC) and field cooling (FC) for the as-intercalated and post-annealed (100 - 250°C, 60 h) samples consisting of Li$_x$(C$_6$H$_{16}$N$_2$)$_y$Fe$_{2-z}$Se$_2$ and FeSe.

As for the reason why the sample post-annealed at 100°C shows $T_c \sim 41$ K in spite of the absence of Phase III in this sample, there might exist a few regions of Phase III, which cannot be detected by the powder x-ray diffraction, in the light of the broad superconducting transition around 41 K. Since the sample including only Phase II has not been obtained, the $T_c$ of Phase II remains unclear. Only the superconducting transition of FeSe is observed in the sample post-annealed at 250°C, which is reasonable taking into account the powder x-ray diffraction result. The decrease in $T_c$ of the sample post-annealed at 200°C is probably due to the disorder of the conducting layer on account of the progress of the deintercalation of HMDA by the post-annealing.

Figure 5 shows the temperature dependence of $\rho$ for the sintered (150°C, 12 h) pellet sample consisting of $Li_x(C_6H_{16}N_2)_yFe_{2-z}Se_2$ and FeSe. Compared with the result of $\rho$ for the pellet sample sintered (190°C, 15 h) in an evacuated glass tube described in our previous paper,[20] the value of $\rho$ is small and zero-resistivity is observed below $\sim 10$ K. These may be attributed to the reduction of insulating HMDA adhering to the grain boundaries by the present sintering of the pellet sample in vacuum ($\sim 10^{-1}$ Pa) using an oil-rotary pump. However, the semi-conducting behavior of $\rho$ around 50 K and the broad superconducting transition seem to be due to the disorder in the progress of the deintercalation of HMDA by the sintering at 150°C for only 12 h. Here, it is noted that the onset temperature of superconductivity, $T_c^{onset}$, is $\sim 48$ K and higher than $T_c$'s of bulk superconductivity of the as-intercalated sample and the sample post-annealed at 150°C for 60 h obtained in the $\chi$ measurements. This is probably owing to the large superconducting fluctuation based on the two-dimensional electronic structure as observed in ethylenediamine-intercalated $A_x(C_2H_8N_2)_yFe_{2-z}Se_2$.[22]

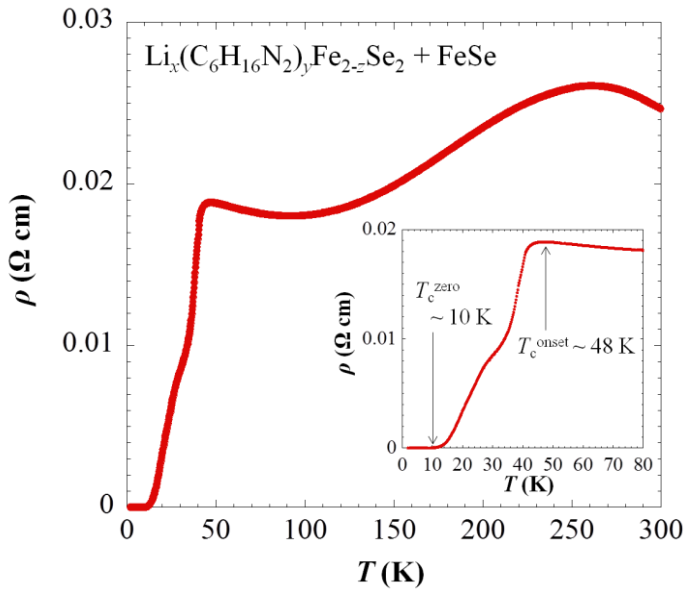

Fig. 5. (Color online) Temperature dependence of the electrical resistivity, $\rho$, for the sintered (150°C, 12 h) pellet sample consisting of $Li_x(C_6H_{16}N_2)_yFe_{2-z}Se_2$ and FeSe. The inset shows the temperature dependence of $\rho$ around $T_c$.

$T_c$'s of the FeSe-based superconductors obtained so far at ambient pressure are summarized in Fig. 6 as a function of $d$.[1,7,8,10,11,13-15,17,19,20,23-27] It is found that $T_c$ tends to increase monotonically with increasing $d$, to be saturated at ~ 45 K and finally to decrease. The present result that $T_c$ of $Li_x(C_6H_{16}N_2)_yFe_{2-z}Se_2$ increases with a decrease in $d$ supports this domic relation between $T_c$ and $d$ in the FeSe-based intercalation superconductors. The increase in $T_c$ with increasing $d$ at $d \leq 9$ Å is probably due to the enhancement of the two-dimensionality in the electronic structure,[28] because the two-dimensionality generally increases the density of states at the Fermi level and moreover improve the nesting condition at the Fermi surface which is important for the spin-fluctuation-mediated superconductivity.

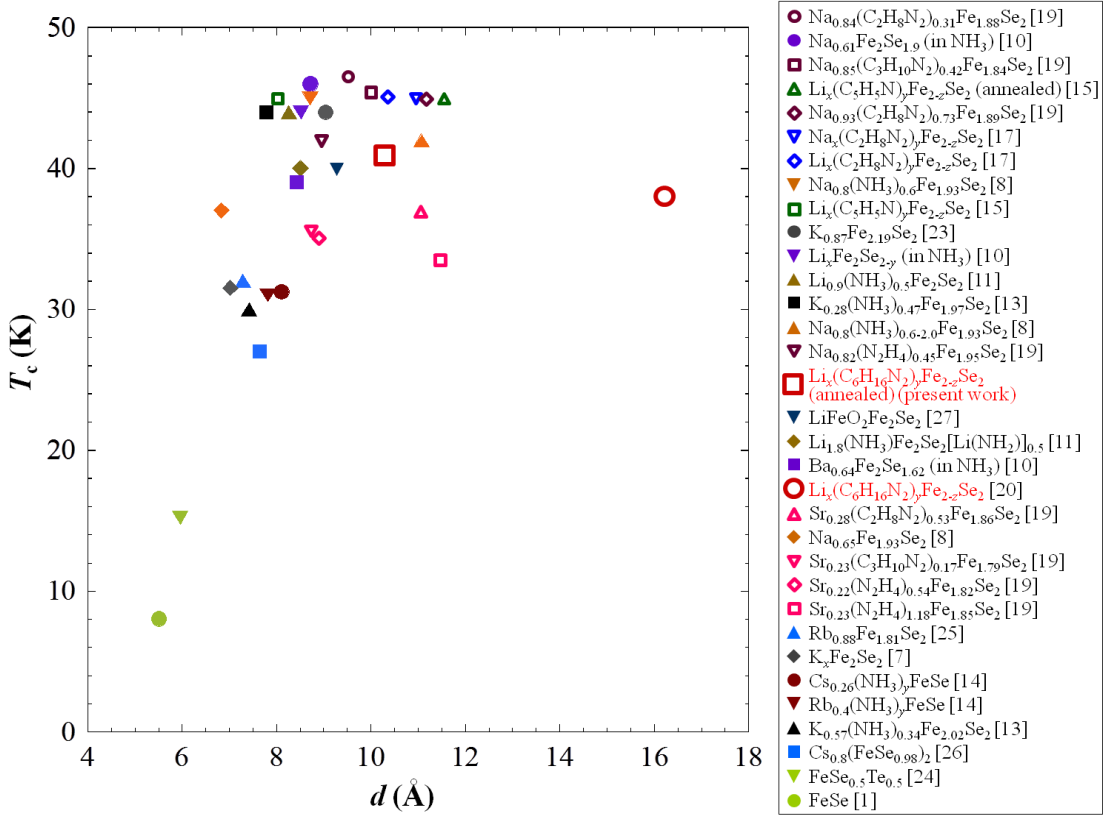

Fig. 6. (Color online) Relation between $T_c$ and the interlayer spacing between neighboring Fe layers, $d$, in the FeSe-based intercalation superconductors.

As for the origin of the decrease in $T_c$ at $d > 9$ Å, it may be due to the decrease of the interlayer coupling of cooper pairs. Considering the superconductivity above 100 K in single-layer FeSe films as mentioned above,[6] however, this may be incorrect. According to the prediction by Guterding *et al*.,[18] the higher $T_c$ of $Li_x(C_6H_{16}N_2)_yFe_{2-z}Se_2$ post-annealed at 150°C for 60 h than that of the as-intercalated sample may be interpreted as being due to the electron doping into the FeSe layer caused by the deintercalation of part of intercalated HMDA, because this is likely when part of intercalated Li$^+$ ions make bonds with possible hydrogen-deficient HMDA molecules and there remain intercalated Li$^+$ ions in spite of the post-annealing. Finally, it is noted that the domic relation between $T_c$ and $d$ may be understood in terms of the empirical relation between $T_c$ and the anion height from the Fe layer in the iron-based superconductors.[29] To confirm it, the detailed analysis of the crystal structures of the FeSe-based intercalation superconductors is necessary.

## 4. Summary

In summary, we have investigated post-annealing effects on the crystal structure and superconductivity in Li$_x$(C$_6$H$_{16}$N$_2$)$_y$Fe$_{2-z}$Se$_2$. It has been found with increasing post-annealing temperature that HMDA deintercalates gradually so that the $d$ value decreases discretely from 16.17(3) Å (Phase I) in the as-intercalated sample to ~ 14 Å (Phase II) and ~ 10 Å (Phase III), indicating the appearance of phases different from the as-intercalated one. Finally, HMDA has been found to perfectly deintercalate at a high temperature above ~ 250°C. The two steps of the reduction of $d$ through the post-annealing have been speculated as being due to stable inclinations of HMDA intercalated between FeSe layers. It has been found that $T_c$ is 41 K in Phase III, namely, in the sample post-annealed at 150°C of Li$_x$(C$_6$H$_{16}$N$_2$)$_y$Fe$_{2-z}$Se$_2$ with $d$ = 10.30(2) Å, while $T_c$ is 38 K in Phase I of the as-intercalated sample with $d$ = 16.17(3) Å. This result supports the domic relation between $T_c$ and $d$ in the FeSe-based intercalation superconductors. As to the reason why $T_c$ is saturated and decreases at $d$ > 9 Å in the FeSe-based intercalation superconductors, it might be due to the decrease of the interlayer coupling of cooper pairs, the possible decrease of electron carriers in the FeSe layer due to the increase of the amount of organic molecules or the possible change of the anion height from the Fe layer. In addition, both zero-resistivity below ~ 10 K and large superconducting fluctuation above $T_c$ have been observed in the $\rho$ measurements of the sintered (150°C, 12 h) pellet sample of Li$_x$(C$_6$H$_{16}$N$_2$)$_y$Fe$_{2-z}$Se$_2$.

**Acknowledgements**

We would like to thank M. Ishikuro of Institute for Materials Research, Tohoku University, for his aid in the ICP-OES analysis.

**References**

1) F. -C. Hsu, J. -Y. Luo, K. -W. Yeh, T. -K. Chen, T. -W. Huang, P. M. Wu, Y. -C. Lee, Y. -L. Huang, Y. -Y. Chu, D. -C. Yan, and M. -K. Wu, Proc. Natl. Acad. Sci. **105**, 14262 (2008).
2) S. Medvedev, T. M. McQueen, I. A. Troyan, T. Palasyuk, M. I. Eremets, R. J. Cava, S. Naghavi, F. Casper, V. Ksenofontov, G.Wortmann and C. Felser, Nat. Mater. **8**, 630 (2009).
3) S. Margadonna, Y. Takabayashi, Y. Ohishi, Y. Mizuguchi, Y. Takano, T. Kagayama. T. Nakagawa, M. Takata, and K. Prassides, Phys. Rev. B **80**, 064506 (2009).
4) Q. -Y. Wang, Z. Li, W. -H. Zhang, Z. -C. Zhang, J. -S. Zhang, W. Li, H. Ding, Y. -B. Ou, P. Deng, K. Chang, J. Wen, C. -L. Song, K. He, J. -F. Jia, S. -H. Ji, Y. -Y. Wang, L. -L. Wang, X. Chen, X. -C. Ma and Q. -K. Xue, Chin. Phys. Lett. **29**, 037402 (2012).
5) S. Tan, Y. Zhang, M. Xia, Z. Ye, F. Chen, X. Xie, R. Peng, D. Xu, Q. Fan, H. Xu, J. Jiang, T. Zhang, X. Lai, T. Xiang, J. Hu, B. Xie, and D. Feng, Nat Mater. **12**, 634 (2013).
6) J. -F. Ge, Z. -L. Liu, C. Liu, C. -L. Gao, D. Qian, Q. -K. Xue, Y. Liu, and J. -F. Jia, Nat. Mater. **14**, 285 (2015).
7) J. Guo, S. Jin, G. Wang, S. Wang, K. Zhu, T. Zhou, M. He, and X. Chen, Phys. Rev. B **82**, 180520(R) (2010).
8) J. Guo, H. Lei, F. Hayashi and H. Hosono, Nat. Commun. **5**, 4756 (2014).
9) H. Abe, T. Noji, M. Kato, and Y. Koike, Physica C **470**, S487 (2010).
10) T. P. Ying, X. L. Chen, G. Wang, S. F. Jin, T. T. Zhou, X. F. Lai, H. Zhang, and W. Y. Wang, Sci. Rep. **2**, 426 (2012).
11) E. -W. Scheidt, V. R. Hathwar, D. Schmitz, A. Dunbar, W. Scherer, F. Mayr, V. Tsurkan, J. Deisenhofer, and A. Loidl, Eur. Phys. J. B **85,** 279 (2012).
12) M. Burrard-Lucas, D. G. Free, S. J. Sedlmaier, J. D. Wright, S. J. Cassidy, Y. Hara, A. J. Corkett, T. Lancaster, P. J. Baker, S. J. Blundell, and S. J. Clarke, Nature Mater. **12**, 15 (2013).
13) T. Ying, X. Chen, G. Wang, S. Jin, X. Lai, T. Zhou, H. Zhang, S. Shen, and W.Wang, J. Am. Chem. Soc. **135,** 2951 (2013).
14) L. Zheng, M. Izumi, Y. Sakai, R. Eguchi, H. Goto, Y. Takabayashi, T. Kambe, T. Onji, S. Araki, T. C. Kobayashi, J. Kim, A. Fujiwara, and Y.Kubozono, Phys. Rev. B **88**, 094521 (2013).
15) A. Krzton-Maziopa, E. V. Pomjakushina, V. Y. Pomjakushin, F. von Rohr, A. Schilling, and K. Conder, J. Phys.: Condens. Matter **24**, 382202 (2012).

16) T. Hatakeda, T. Noji, T. Kawamata, M. Kato, and Y. Koike, J. Phys. Soc. Jpn. **82**, 123705 (2013).

17) T. Noji, T. Hatakeda, S. Hosono, T. Kawamata, M. Kato, and Y. Koike, Physica C **504**, 8 (2014).

18) D. Guterding, H. O. Jeschke, P. J. Hirschfeld and R. Valenti, Phys. Rev. B **91**, 041112(R) (2015).

19) F. Hayashi, H. Lei, J. Guo and H. Hosono, Inorg. Chem. **54**, 3346 (2015).

20) S. Hosono, T. Noji, T. Hatakeda, T. Kawamata, M. Kato, and Y. Koike, J. Phys. Soc. Jpn. **83**, 113704 (2014).

21) F. Izumi, and K. Momma, Solid State Phenom. **130**, 15 (2007).

22) T. Hatakeda, T. Noji, S. Hosono, T. Kawamata, M. Kato, and Y. Koike, J. Phys.: Conf. Ser. **568**, 022032 (2014).

23) A. -M. Zhang, T. -L. Xia, K. Liu, W. Tong, Z. -R. Yang, and Q. -M. Zhang, Sci. Rep. **3**, 1216 (2013).

24) K. -W. Yeh, T. -W. Huang, Y. -L. Huang, T. -K. Chen, F. -C. Hsu, P. M. Wu, Y. -C. Lee, Y. -Y. Chu, C. -L. Chen, J. -Y. Luo, D. -C. Yan, and M. -K. Wu, Europhys. Lett. **84**, 37002 (2008).

25) A. F. Wang, J. J. Ying, Y. J. Yan, R. H. Liu, X. G. Luo, Z. Y. Li, X. F. Wang, M. Zhang, G. J. Ye, P. Cheng, Z. J. Xiang, and X. H. Chen, Phys. Rev. B **83,** 060512(R) (2011).

26) A. Krzton-Maziopa, Z. Shermadini, E. Pomjakushina, V. Pomjakushin, M. Bendele, A. Amato, R. Khasanov, H. Luetkens, and K. Conder, J. Phys.: Condens. Matter **23**, 052203 (2011).

27) X. F. Lu, N. Z. Wang, G. H. Zhang, X. G. Luo, Z. M. Ma, B. Lei, F. Q. Huang and X. H. Chen, Phys. Rev. B **89**, 020507(R) (2013).

28) T. Takano, T. Kishiume, Y. Taguchi, and Y. Iwasa, Phys. Rev. Lett. **100**, 247005 (2008).

29) Y. Mizuguchi, Y. Hara, K. Deguchi, S. Tsuda, T. Yamaguchi, K. Takeda, H. Kotegawa, H. Tou, and Y. Takano, Supercond. Sci. Technol. **23**, 054013 (2010).